\documentclass[12pt]{article}
\usepackage{amssymb,amsmath,cite}
\catcode`\@=11

\@addtoreset{equation}{section}
\def\theequation{\arabic{section}.\arabic{equation}}
     
\catcode`\@=11
\def\thesection{\arabic{section}}

\def\appendix{\setcounter{section}{0}
        \def\thesection{Appendix.}
        \def\theequation{\Alph{section}.\arabic{equation}}}
\def\section{\@startsection{section}{1}{\z@}{3.5ex plus 1ex minus
   .2ex}{2.3ex plus .2ex}{\large\bf}}
\def\subsection{\@startsection{subsection}{1}{\z@}{3.5ex plus 1ex minus
   .2ex}{2.3ex plus .2ex}{\bf}}

\long\def\@makefntext#1{\parindent 0cm\noindent
\hbox to 1em{\hss$^{\@thefnmark}$}#1}

\def\IR{{\hbox{{\rm I}\kern-.2em\hbox{\rm R}}}}
\def\IH{{\hbox{{\rm I}\kern-.2em\hbox{\rm H}}}}
\def\IC{{\ \hbox{{\rm I}\kern-.6em\hbox{\bf C}}}}
\def\IZ{{\hbox{{\rm Z}\kern-.4em\hbox{\rm Z}}}}

\def\tg{{\tilde g}}
\def\th{{\tilde h}}
\newcommand{\beq}{\begin{equation}}
\newcommand{\eeq}{\end{equation}}

\begin{document}
\begin{titlepage}
\vspace{.5in}
\begin{flushright}
UCD-04-19\\
gr-qc/0403060\\
March 2004\\
(revised June 2004)\\
\end{flushright}
\vspace{.5in}
\begin{center}
{\large\bf
 Model-Dependence of Shapiro Time Delay\\[1ex] 
 and the
 ``Speed of Gravity/Speed of Light'' Controversy}\\
\vspace{.4in}
{S.~C{\sc arlip}\footnote{\it email: carlip@physics.ucdavis.edu}\\
       {\small\it Department of Physics}\\
       {\small\it University of California}\\
       {\small\it Davis, CA 95616}\\{\small\it USA}}
\end{center}

\vspace{.5in}
\begin{center}
{\large\bf Abstract}
\end{center}
\begin{center}
\begin{minipage}{4.75in}
{\small Fomalont and Kopeikin have recently
succeeded in measuring the velocity-dependent component of the
Shapiro time delay of light from a quasar passing behind Jupiter.
While there is general agreement that this observation tests the 
gravitomagnetic properties of the gravitational field,
a controversy has emerged over the question of whether 
the results depend on the speed of light, $c$, or the speed of gravity, 
$c_g$.  By analyzing the Shapiro time delay in a set of ``preferred 
frame'' models, I demonstrate that this question is ill-posed: the 
distinction can only be made in the context of a class of theories 
in which $c\ne c_g$, and the answer then depends on the specific 
class of theories one chooses.  It remains true, however, that for 
a large class of theories ``close enough'' to general relativity, 
the \emph{leading} contribution to the time delay depends on $c$ 
and not $c_g$; within this class, observations are thus not yet 
accurate enough to measure the speed of gravity.
}
\end{minipage}
\end{center}
\end{titlepage}
\addtocounter{footnote}{-1}

\section{Introduction}

In a remarkable experiment, Fomalont and Kopeikin \cite{Fomalont}
recently observed the Shapiro time delay of light from the quasar 
QSO J0842+1835 as its image passed within 3.7 arcminutes of Jupiter.  
Using VLBI, they succeeded in measuring not only the first-order
effects, but the dependence on Jupiter's velocity as well.  The
results agree well with the predictions of general relativity, and
there is general agreement that the observation has tested the 
gravitomagnetic effect of Jupiter's motion.  Within the framework
of general relativity, this is essentially equivalent to a confirmation
of the Lorentz transformation properties of the gravitational field.
A controversy has emerged, however, over the question of whether the
results depend on the speed of gravity, $c_g$---in which case the
observation can be construed as a measurement of $c_g$ as well---or 
only on the speed of light, $c$ 
\cite{Asada,Will,Kopeikin0,Kopeikin,Kopeikin2,Kopeikin3,Faber,Frittelli}.  

More precisely, the delay predicted by standard general relativity 
due to a gravitating body of mass $m_a$ and velocity ${\bf v}_a$ 
takes the form
\begin{align}
\Delta =& -\frac{2Gm}{{\tilde c}^3}\left( 1 - \frac{{\bf\widehat k}\cdot{\bf v}}%
  {{\tilde c}}\right)\ln \left( r_a - {\bf K}\cdot{\bf r}_a\right)
\nonumber\\
&\hbox{with}\ \ {\bf K} = {\bf\widehat k} - \frac{1}{\tilde c}{\bf\widehat k}%
  \times({\bf v}_a\times{\bf\widehat k}) ,
\label{a1}
\end{align}
where ${\bf r}_a$ is the separation vector between the gravitating
body and the observer and ${\bf\widehat k}$ is the unit vector in the
direction of the incoming light ray.  The question at issue is whether
$\tilde c$ in eqn.\ (\ref{a1}) should be understood as $c$ or $c_g$,
or some combination---perhaps even a different combination in each of
its appearances.

Let us begin with two simple observations.  First, within standard
general relativity, $c$ and $c_g$ are identical: gravity propagates
along null geodesics \cite{Low}, and $c$ and $c_g$ are determined by
exactly the same metric information.  This equality is deeply embedded 
in the Einstein field equations, since, as is well known, the field
equations themselves determine the motion of the sources \cite{Damour}.
One thus cannot simply declare $c$ and $c_g$ to be two separate
parameters in general relativity; to distinguish the two speeds, 
one must instead look at a broader class of ``comparison theories''
in which $c\ne c_g$ to determine the separate dependencies of the
time delay on the two speeds.  Second, it is not obvious that the 
answer will be independent of the class of ``comparison theories''
one chooses. Indeed, Will's PPN formalism \cite{Will} and Kopeikin's
alternative PPN formalism \cite{Kopeikin0,Kopeikin} may be seen in 
part as two different choices that yield two different results.

In this paper, I evaluate the Shapiro time delay in a class of 
preferred frame theories described by Jacobson and Mattingly \cite{Jac}
(see also \cite{Kost}).  While these are by no means the most general
models with $c\ne c_g$, they are sufficiently general to clearly show
that the parameter $\tilde c$ in (\ref{a1}) is model-dependent, and
can depend in complicated ways on $c$ and $c_g$.  The debate over 
whether Fomalont and Kopeikin have measured the speed of gravity is 
thus misguided: the interpretation of their measurement depends on 
a choice of ``comparison theory,'' and cannot be narrowed down unless 
that choice is strongly constrained by other means.  

I also show, however, that for theories in the class I consider---and 
probably for a considerably more general class of theories that are 
sufficiently ``close'' to general relativity---the factor of $\tilde c$
that appears inside the logarithm of (\ref{a1}) is the speed of light.  
Since for ordinary general relativity this term gives the dominant
contribution to the Fomalont-Kopeikin observation, it is arguably true 
that observations are not yet sensitive enough to determine $c_g$.

\section{A Class of Preferred Frame Theories \label{S2}}

Suppose we have a theory in which light and gravity propagate at 
fixed speeds, but in which $c\ne c_g$.  Such a theory will have 
two sets of ``null cones,'' one determined by the propagation of 
gravity and one by the propagation of light, and thus two metrics.%
\footnote{There are actually a number of independent ways in which
the parameter $c$ can enter; see \cite{Uzan}.  I consider only the
simplest.}  It will also violate Lorentz invariance, in the sense 
that it will admit a preferred frame, the frame in which the speed 
of gravity and the speed of light are both isotropic.  Such violations 
are strongly constrained by experiment, but the strongest constraints 
are nongravitational \cite{Lor}; as long as light and matter see 
the same metric, there is room for the gravitational metric to be 
different.

The simplest, although certainly not the only, way to build such a 
theory is to start with a gravitational metric $g_{ab}$ and a unit 
timelike one-form field $u_a$, and to introduce a second metric $\tg_{ab}$ 
to describe the propagation of light and matter, 
\beq
\tg_{ab} = g_{ab} - \epsilon u_au_b , \quad g^{ab}u_au_b=1 ,
\label{b1}
\eeq
where $\epsilon$ is a fixed parameter.  To keep notation simple, indices 
will be raised and lowered by $g_{ab}$ only; in particular, $\tg^{ab} 
= g^{ac}\tg_{cd}g^{bd}$ will not be the inverse of $\tg_{ab}$.  I choose 
coordinates such that $g_{ab}$ reduces to the standard Minkowski metric 
$\eta_{ab}$ in flat spacetime; in the preferred frame $u_i=0$, we then 
have $\tg_{ab} = \hbox{diag}(1-\epsilon,-1,-1,-1)$.  The speed of light 
is thus
\beq
c = \sqrt{1-\epsilon}
\label{b2}
\eeq
in units $c_g=1$.  If $c>c_g$, the absence of gravitational Cerenkov 
radiation places extremely strong limits on the difference \cite{Moore};
I therefore take $\epsilon$ to be positive.

Let us now consider a theory with three dynamical sectors: gravity, the 
vector field $u$, and matter (including light):\\

\noindent {\bf 1.\ Gravity:} I take the gravitational action to be
the usual Einstein-Hilbert action for the metric $g_{ab}$,
\beq
I_{\mathit{grav}} = \frac{1}{2\kappa^2}\int d^4x\sqrt{-g}\,R[g]
\label{b3}
\eeq
In standard general relativity, the coupling constant $\kappa$ depends
on $c$; here, that dependence will be determined below from the Newtonian
limit.\\

\noindent {\bf 2.\ Vector:} The vector field $u_a$ has four possible
kinetic terms quadratic in derivatives \cite{Jac2}.  I choose the 
simplest one, that of \cite{Jac}:
\beq
I_u = \frac{\beta}{2\kappa^2}\int d^4x\sqrt{-g}\left[ \nabla_au_b%
  (\nabla^au^b - \nabla^bu^a) + \lambda(g^{ab}u_au_b - 1) \right].
\label{b4}
\eeq
With this choice the longitudinal mode becomes nonpropagating, and 
the propagating modes of $u$ and $g$ all travel at speed $c_g$; thus 
the ``speed of gravity'' is unique and well-defined \cite{Jac2}.  
The Lagrange multiplier $\lambda$ ensures that $u$ remains a unit 
timelike vector.  The coupling constant $\beta$ is, for now, arbitrary; 
we shall see later that a particularly interesting case is that in 
which it is of order $\epsilon$.\\

\noindent {\bf 3.\ Matter:} As noted above, nongravitational
observations strongly suggest that all forms of matter see very
nearly the same ``speed of light.''  I therefore take a generic
matter action to be of the form
\beq
I_{\mathit{mat}} = I_{\mathit{mat}}[\psi,\tg]
\label{b5}
\eeq
with the standard minimal coupling to $\tg$.  The definition of the
stress-energy tensor in a bimetric theory is ambiguous; I choose
the convention
\beq
\delta I_{\mathit{mat}} = \frac{1}{2}\int d^4x \sqrt{-g}\,T^{ab}\delta\tg_{ab}
  = \frac{1}{2}\int d^4x \sqrt{-g}\left( T^{ab}\delta g_{ab} - 2\epsilon
  T^{ab}u_b\delta u_a\right) ,
\label{b6}
\eeq
where the second equality comes from the fact that $I_{\mathit{mat}}$
depends on $g$ and $u$ only through the combination $\tg$.  Note the 
choice of $\sqrt{-g}$ rather than $\sqrt{-\tg}$ in the integral.  A 
different choice would introduce extra factors of $c/c_g$.  Once the
final results are expressed in terms of physically measured quantities,
however, this ambiguity will disappear.

Two special cases will be important.  The first is that of a point 
particle of mass $m$ moving along a world line $\gamma$,
\beq
I_m = m\int_\gamma d{\tilde s} = m\int d^4x \int d\lambda\,
  {\bar\delta}^4(x-z(\lambda)) \left(\tg_{ab}\frac{dx^a}{d\lambda}
  \frac{dx^b}{d\lambda}\right)^{1/2} ,
\label{b7}
\eeq
where $\bar\delta$ is a ``densitized'' delta function.  An easy
computation gives
\beq
T_m^{ab} 
= m\int_\gamma ds\, \delta^4(x-z(s))(1-\epsilon(u\cdot v)^2)^{-1/2}v^av^b
\qquad \hbox{with}\ \ v^a=\frac{dx^a}{ds}
\label{b8}
\eeq
where $\delta$ is the ordinary (``undensitized'') delta function, i.e.,
$\int d^4x\sqrt{-g}\,\delta^4(x)=1$.

The second special case is light, which we can describe by the standard
Maxwell action coupled to the metric $\tg$,
\beq
I_{\mathit{EM}} = \frac{1}{4}\int d^4x\sqrt{-\tg}\,\tg^{ac}\tg^{bd}
  (\partial_aA_b - \partial_bA_a)(\partial_cA_d - \partial_dA_c) .
\label{b9}
\eeq
By standard geometric optics arguments (see, for example, \cite{MTW}),
it follows that light will follow the null geodesics of the metric
$\tg$.  To compute Shapiro time delay, we will thus need to understand
the dynamics of this metric.

\section{Field Equations and the Weak Field Approximation \label{S3}}

The field equations are easily obtained from eqns.\ (\ref{b3}--\ref{b6}):
\beq
G^{ab} = \kappa^2\left(T_m^{ab} + T_u^{ab}\right) - \beta\lambda u^au^b
\label{c1}
\eeq
\beq
\nabla_a\left(\nabla^au^b - \nabla^bu^a\right) = 
  -\frac{\kappa^2}{\beta}\epsilon\, T_m^{ab}u_a + \lambda u^b
\label{c2}
\eeq
\beq
g^{ab}u_au_b = 1
\label{c3}
\eeq
where
\beq
T_u^{ab} = -\frac{\beta}{\kappa^2}\left( H^a{}_cH^{bc} - \frac{1}{4}g^{ab}
  H_{cd}H^{cd}\right) \qquad \hbox{with}\ H^{ab} = \nabla^au^b - \nabla^bu^a .
\label{c4}
\eeq
Differentiating (\ref{c1}) and (\ref{c2}) and using the identities 
$\nabla_a G^{ab} =0$ and $\nabla_a\nabla_b H^{ab}=0$, we obtain a
conservation law
\beq
\nabla_a T_m^{ab} = \epsilon\nabla_a\left(T_m^{ac}u_cu^b\right)
  - \epsilon T_m^{ac}u_a\nabla^bu_c .
\label{c5}
\eeq
It is not hard to show that this implies that ${\widetilde\nabla}_aT_m^{ab}=0$,
where $\widetilde\nabla$ is the covariant derivative compatible with the metric
$\tg$.  This means, in particular, that the field equations are compatible
with the matter equations of motion in the geometry determined by $\tg$.

\subsection{Weak field approximation}
We next need the weak field approximation of these field equations.  Let
\beq
g_{ab} = \eta_{ab} + h_{ab}, \qquad u_a = {\bar u}_a + w_a
\label{c6}
\eeq
where $\bar u$ is a constant timelike one-form such that $\eta^{ab}{\bar u}_a
{\bar u}_b = 1$, and where $h$ and $w$ are small.
Define
\beq
\chi^b = \partial_a\left(h^{ab} - \frac{1}{2}\eta^{ab}h\right) ,
\label{c7}
\eeq
where, as usual in the weak field approximation, we now raise and lower
indices with the flat metric $\eta$.  It is then straightforward to check
that to lowest order, the field equations become
\beq
\Box h_{ab} - \partial_a\chi_b - \partial_b\chi_a = -2\kappa^2
  \left(T_{ab} - \frac{1}{2}\eta_{ab}T\right) + 2\beta\lambda
  \left({\bar u}_a{\bar u}_b - \frac{1}{2}\eta_{ab}\right)
\label{c8}
\eeq
\beq
\Box w_b = \partial_b(\partial_aw^a) - \frac{\kappa^2}{\beta}
  \epsilon T_{bc}{\bar u}^c + \lambda{\bar u}_b 
\label{c9}
\eeq
\beq
2{\bar u}_aw^a - h^{ab}{\bar u}_a{\bar u}_b = 0 ,
\label{c10}
\eeq
where $T^{ab}$ is now the matter metric alone ($T_u^{ab}$ vanishes
at this order).

The ``matter'' metric $\tg$ can similarly be expanded:
\beq
\tg_{ab} = {\tilde\eta}_{ab} + \th_{ab} \approx 
  (\eta_{ab} - \epsilon{\bar u}_a{\bar u}_b) + 
  (h_{ab} - \epsilon{\bar u}_aw_b - \epsilon {\bar u}_bw_a) .
\label{c11}
\eeq
Combining (\ref{c8}) and (\ref{c9}) and choosing a gauge
\beq
{\tilde\chi}_a = \chi_a - \epsilon{\bar u}_a\partial_bw^b = 0 ,
\label{c12}
\eeq
we find that
\beq
\Box\th_{ab} = -2\kappa^2 \left( T_{ab} - \frac{1}{2}\eta_{ab}T
  - \frac{\epsilon^2}{2\beta}\left( {\bar u}_aT_b{}^c{\bar u}_c
  + {\bar u}_bT_a{}^c{\bar u}_c \right)\right) + 2(\beta-\epsilon)
  \lambda{\bar u}_a{\bar u}_b - \beta\lambda\eta_{ab} .
\label{c13}
\eeq
Note that if $\beta=\epsilon\ll 1$, this has nearly the same form 
as Kopeikin's modified field equations, eqns.\ (2.9) and (2.16) of 
\cite{Kopeikin}.
  
To proceed further, we also need to understand the dynamics
of $\lambda$.  Note first from (\ref{c2}) that $\lambda$ is
small (of order $w$).  Taking the divergence of (\ref{c2}),
we then see that to the order at which we are working,
\beq
{\bar u}^b\nabla_b\lambda=0 ,
\label{c14}
\eeq
i.e., $\lambda$ is nonpropagating \cite{Jac2}.  More precisely, 
choose a coordinate system ${\bar u} = (1,0,0,0)$, so that by 
(\ref{c10}), $w^0 = \frac{1}{2}h^{00}$.  Decompose $w^i$ into 
a transverse component $w^{iT}$ ($\partial_iw^{iT}=0$) and a 
longitudinal component $\partial^i\omega$.  Then (\ref{c9}) 
yields
\beq
\lambda = {\bf\nabla}^2(\partial_0\omega - \frac{1}{2}h_{00}) 
  + \frac{\kappa^2}{\beta}\epsilon T_{00}
\label{c15}
\eeq
with no further equation for $\omega$.  We thus can---and for the
remainder of this paper, will---choose initial data for $\omega$
such that $\lambda=0$ at the required order.  

(One might worry that such data are unstable, since for a static 
source, (\ref{c15}) implies that $\omega$ grows linearly in
time.  This instability is an artifact of our approximation, 
however, and disappears at higher orders \cite{Jac}.  Note that 
in the vacuum case, $\lambda=0$ for the analog of the Schwarzschild 
metric \cite{Jac}.)

\subsection{Newtonian limit}
Let us next consider the Newtonian limit.  Since matter moves
along geodesics with respect to the metric $\tg$, this limit takes
the usual form
\beq
\frac{d^2x^i}{dt^2}\approx \frac{1}{2}\tg^{ij}\partial_j\th_{00} 
  \approx -\frac{1}{2}\partial_i\th_{00} ,
\label{c16}
\eeq
where the last approximation requires that the velocity $u^i$
with respect to the preferred frame be much less than $c$, and
therefore negligible in this limit.  For the correct Newtonian
limit, we thus require that $\th_{00} = 2\Phi$, where $\Phi$ is
the Newtonian potential.  By (\ref{b8}) and (\ref{c13}), on the 
other hand,
\beq
{\bf\nabla}^2\th_{00} \approx 2\kappa^2\left( T_{00} - \frac{1}{2}T
  - \frac{\epsilon^2}{\beta}T_{00} \right) \approx
  \kappa^2\left( 1 - \frac{2\epsilon^2}{\beta}\right)(1-\epsilon)^{-1/2}
  m\,\delta^3(x-z) .
\label{c17}
\eeq
We thus find that
\beq
\kappa^2 = 8\pi G(1-\epsilon)^{1/2}
  \left( 1 - \frac{2\epsilon^2}{\beta}\right)^{-1} .
\label{c18}
\eeq

\section{Shapiro Time Delay}

We are at last in a position to compute the Shapiro time delay for this
class of theories.  We follow the basic procedure of \cite{Will}.  In
the absence of gravity, light follows a straight line.  In the presence
of a weak field, a light path can therefore be written as
\beq
x^i = x^i_e + k^i(t-t_e) + y^i(t) 
\label{d1}
\eeq
with $y^i$ small.  (The suffix $e$ denotes the point of emission.)  Let
\beq 
N^a = \frac{dx^a}{dt} = k^a + \frac{dy^a}{dt}
\label{d2}
\eeq
be the tangent vector to the light's trajectory, with $k^0=1$ and $y^0=0$.  
As noted in section \ref{S2}, this vector must be null with respect to the 
metric $\tg_{ab} = {\tilde\eta}_{ab} + \th_{ab}$.  Thus to lowest order
\beq
(\eta_{ab} - \epsilon{\bar u_a}{\bar u}_b)k^ak^b = 0 = 
  1 - |{\bf k}|^2 - \epsilon({\bar u_0} - {\bf k}\cdot{\bf\bar u})^2
  \approx (1-\epsilon) - |{\bf k}|^2 + 2\epsilon{\bf k}\cdot{\bf\bar u} ,
\label{d3}
\eeq
where in the last line I have dropped terms of order $|{\bf\bar u}|^2$.  
Not surprisingly, the speed of light is anisotropic: the speed in 
the $\bf\widehat k$ direction is
\beq
c_{\bf k} = \left( 1-\epsilon + 2\epsilon{\bf k}\cdot{\bf\bar u}\right)^{1/2}
  \approx c + \epsilon\frac{{\bf k}\cdot{\bf\bar u}}{c} .
\label{d4}
\eeq
At the next order, 
\beq
2(\eta_{ab}-\epsilon{\bar u_a}{\bar u}_b)k^a\frac{dy^b}{dt} 
  + \th_{ab}k^ak^b = 0 ,
\label{d5}
\eeq
implying, again up to terms of order $|{\bf\bar u}|^2$, that
\beq
\frac{d\ }{dt}({\bf k}\cdot{\bf y}) = 
  -\epsilon\frac{d\ }{dt}({\bf\bar u}\cdot{\bf y}) + \frac{1}{2}\th_{ab}k^ak^b .
\label{d6}
\eeq

Now consider the square of (\ref{d1}):
\beq
\left| {\bf x} - {\bf x}_e \right|^2 \approx c_{\bf k}^2(t-t_e)^2 + 
  2{\bf k}\cdot{\bf y}(t-t_e) ,
\label{d7}
\eeq
or, using (\ref{d6}),
\begin{align}
t-t_e \approx \frac{|{\bf x} - {\bf x}_e|}{c_{\bf k}} - 
  \frac{{\bf k}\cdot{\bf y}}{c_{\bf k}^2} &=
  \frac{|{\bf x} - {\bf x}_e|}{c_{\bf k}} - \frac{1}{c_{\bf k}^2}
  \int_{t_e}^t dt\left[\frac{1}{2}\th_{ab}k^ak^b\right] 
  + \epsilon\frac{{\bf\bar u}\cdot{\bf y}(t)}{c_{\bf k}^2} \nonumber\\
  &= \frac{|{\bf x} - {\bf x}_e|}{c_{\bf k}} + \Delta(t,t_e) ,
\label{d8}
\end{align}
where to this order the integral is along the unperturbed straight light
path.  For $\epsilon=0$, this reduces to the standard Shapiro time delay
formula (note that my signature conventions differ from \cite{Will}).
For $\epsilon\ne0$, it contains two new features, anisotropic 
``preferred frame'' terms proportional to ${\bf k}\cdot{\bf\bar u}$ and 
$\epsilon$-dependent terms that might distinguish the speed of light
and the speed of gravity.  The former are certainly of interest, and
might be useful for constraining models of this sort, but they are not
to the issue being addressed in this paper.  Therefore, following Will  
\cite{Will} and Kopeikin \cite{Kopeikin}, I will discard these terms, 
by choosing coordinates ${\bar u}^0=1$, ${\bf\bar u} = 0$.

By (\ref{b8}) and (\ref{c13}), the contribution to (\ref{d8}) of 
a point mass $m$ and four-velocity $v^a$ is given by
\begin{align}
\Box(\th_{ab}k^ak^b) &= -2\kappa^2 \left( T_{ab}k^ak^b - \frac{\epsilon}{2}T
  - \frac{\epsilon^2}{\beta}T_{0a}k^a \right)\label{d9}\\
  &= -16\pi Gm \left(1 - \frac{2\epsilon^2}{\beta}\right)^{-1} \int ds\,
  \delta^4(x-z(s)) 
  \left[\left( 1 - \frac{\epsilon}{2} - \frac{\epsilon^2}{\beta}\right)
  - 2\left( 1 - \frac{\epsilon^2}{2\beta}\right){\bf v}\cdot{\bf k}\right] 
  \nonumber
\end{align}
where I have used (\ref{c18}) for $\kappa^2$ and discarded terms of order 
$|{\bf v}|^2$.  The solution takes the standard Li{\'e}nard-Weichert form
\beq
\frac{1}{2}\th_{ab}k^ak^b = (1+\gamma)Gm
  \frac{1 - (2+\zeta)\frac{{\bf v}\cdot{\bf k}}{c^2}}{|{\bf x} - {\bf z}(s_R)|
  - {\bf v}(s_R)\cdot({\bf x} - {\bf z}(s_R))}
\label{d10}
\eeq
where $s_R$ is the retarded time,\footnote{Note that since $\Box$ in
(\ref{d9}) is the gravitational d'Alembertian, this expression depends
on $c_g$, not $c$.}
\beq
s_R = t - |{\bf x} - {\bf z}(s_R)| ,
\label{d11}
\eeq
and
\begin{align}
1 + \gamma &= 2\left(1 - \frac{2\epsilon^2}{\beta}\right)^{-1}
  \left(1 - \frac{\epsilon}{2} - \frac{\epsilon^2}{\beta}\right) \nonumber\\
\zeta &= -2\left[1-(1-\epsilon)\left( 1 - \frac{\epsilon^2}{2\beta}\right)
  \left(1 - \frac{\epsilon}{2} - \frac{\epsilon^2}{\beta}\right)^{-1}\right] .
\label{d12}
\end{align}
I have chosen notation in such a way that this expression is identical
to that of \cite{Will}; one can therefore read off the results directly
from that paper (although one must keep careful track of factors of $c$):
\begin{align}
\Delta(t,t_e) &= - \frac{(1+\gamma)Gm}{c^3}\left\{
  \left[ 1 - (1+\zeta)\frac{{\bf k}\cdot{\bf v}}{c^2}\right]\ln
  \left(r - {\bf K}\cdot{\bf r}\right)\right\}\nonumber\\
&\hbox{with}\quad {\bf K} = {\bf\widehat k} - \frac{1}{c}{\bf\widehat k}%
  \times({\bf v}_a\times{\bf\widehat k}) , \qquad
  {\bf\widehat k} = {\bf k}/c.
\label{d13}
\end{align}
There is one basic physical difference between this expression and the 
results of \cite{Will}, though: while the expression in \cite{Will} had 
no apparent dependence on the speed of gravity, here such a dependence 
is explicit.

Two special cases are of interest.  First, suppose $\epsilon$ is small 
and $\beta$ is of order unity.  Then $1+\gamma\approx 2(1-\epsilon/2)
\approx 2(c/c_g)$ and $1+\zeta \approx 1-\epsilon = (c/c_g)^2$.  The 
first-order time delay thus acquires a prefactor $Gm/c_g{}c^2$, while
the velocity-dependent prefactor is proportional to ${\bf k}\cdot{\bf v}/%
c_g{}^2$. To allow more general possibilities, suppose instead that 
$\beta$ is of order $\epsilon$, say $\beta = \epsilon/b$, with $\epsilon$
small.  Then
\begin{align}
1+\gamma &\approx 2\left( 1 - (1-2b)\frac{\epsilon}{2}\right) 
  \approx 2\left( c/c_g \right)^{1-2b} \nonumber\\
1 + \zeta &\approx \left(1 - (1-b)\epsilon\right)
  \approx \left( c/c_g \right)^{2(1-b)}
\label{d14}
\end{align}
and thus
\beq
\Delta(t,t_e) = - \frac{2Gm}{c_1{}^3}\left\{
  \left[ 1 - \frac{{\bf k}\cdot{\bf v}}{c_2{}^2}\right]\ln
  \left(r - {\bf K}\cdot{\bf r}\right)\right\}
\label{d15}
\eeq
with
\beq
c_1 = c^{\frac{2(1+b)}{3}}c_g{}^{\frac{1-2b}{3}} \qquad
c_2 = c^bc_g{}^{1-b} .
\label{d16}
\eeq

We see that by varying the coupling $\beta$, one can make the 
velocity-dependent prefactor in (\ref{d13}) depend almost 
arbitrarily on $c$ or $c_g$.  The logarithmic term, on the
other hand, depends only on the speed of light $c$.  Since
in ordinary general relativity this term gives the dominant 
contribution to the Fomalont-Kopeikin measurement, a bit more 
detail may be in order.  The logarithm comes from an integral 
of a retarded potential,
\beq
I = \int \frac{dt}{|{\bf x} - {\bf z}(s_R)|
  - \frac{{\bf v}(s_R)\cdot({\bf x} - {\bf z}(s_R))}{c_g}} ,
\label{d17}
\eeq
where I have restored the explicit factor of $c_g$.  The key
observation---explained in \cite{Carlip} for the case of gravity
and in \cite{Feynman} for the case of electromagnetism, and used
by Will \cite{Will}---is that to first order in $v$, the denominator
of this integral is \emph{independent} of $c_g$.  This does not,
of course, mean that gravity does not propagate at the speed $c_g$.
But to first order, the measurable effects of the propagation 
delay are canceled by velocity-dependent terms in the interaction,
and observations are insensitive to the value of $c_g$ in (\ref{d17}).
More precisely, using (\ref{d11}), one easily finds that the 
denominator of (\ref{d17}) is
\begin{align}
|{\bf x} - {\bf z}(s_R)| &- 
  \frac{{\bf v}(s_R)\cdot({\bf x} - {\bf z}(s_R))}{c_g}
  = |{\bf x} - {\bf z}(s_R) - (t-s_R){\bf v}(t)| 
  + {\cal O}({v^2}/{c_g{}^2})\nonumber\\
  &= |{\bf x} - {\bf z}(t)| + {\cal O}({v^2}/{c_g{}^2})
  = |{\bf x}(t_r) - {\bf z}(t_r) + ({\bf k} - {\bf v})(t-t_r)|
  + {\cal O}({v^2}/{c_g{}^2})
\label{d18}
\end{align}
where $t$ is the ``instantaneous'' time and $t_r$ is the time of 
reception of the ray.  Observe that the speed of gravity $c_g$ 
has dropped out of the expression, which depends, at order $v/c_g$, 
only on the unretarded ``instantaneous'' position of the source.  
Thus to this order, the integral (\ref{d17}) \emph{cannot} depend 
on $c_g$.\footnote{This is the underlying reason that Faber's 
computation with $c_g=\infty$ \cite{Faber} agrees with that of 
Will \cite{Will}.}

An explicit evaluation of (\ref{d17}) confirms this conclusion.  
The logarithmic term in (\ref{d13}) contains factors of $c$ not 
because the retarded potential depends on $c_g$---this factor has 
canceled out at order $v/c_g$---but only because the integrand 
depends on $|{\bf k} - {\bf v}|$, and thus on $|{\bf k}|$.  By 
(\ref{d3}), $|{\bf k}| = c $, and by (\ref{d1}) the $c$ appearing 
here is \emph{directly} the speed of light, that is, the distance
traveled by the light ray divided by its travel time.  As a  
straightforward check of this result, one can repeat the same
computation for a rapidly moving massive particle traveling at a 
subluminal speed $w$ near $c$.  It is not hard to show that to 
first order in $v$, the only change in the argument of the
logarithm in (\ref{d13}) is that $v/c$ is replaced by $v/w$.

\section{Conclusion}

In ordinary general relativity, the speed of gravity and the speed
of light are equal.  This is not merely a coincidence; the two
speeds express the same information about the metric structure
of spacetime.  Most of the controversy over the interpretation of
the Fomalont-Kopeikin observation has implicitly assumed that there
is some unique way to separate these speeds, that is, some unique
deformation of general relativity to a $c\ne c_g$ theory.  We have
now seen explicitly that this is not the case: we have constructed
a class of theories that all reduce to general relativity as 
$c\rightarrow c_g$, but in which the Shapiro time delay depends
on different combinations of $c$ and $c_g$ determined by the choice 
of an otherwise arbitrary coupling constant.

In the process, we have reproduced one result of Will \cite{Will}:
the vector ${\bf K}$ in (\ref{d13}) depends only on $c$, essentially
because the $c_g$-dependence of a retarded potential cancels to
lowest order in $v/c$ \cite{Carlip}.  This is a generic feature of
retarded potentials, and it would take a considerably more drastic
shift away from general relativity to change this result.  We have
also, however, reproduced a result of Kopeikin \cite{Kopeikin0,%
Kopeikin}: if $c\ne c_g$, conservation requires a modification of 
the stress-energy tensor, and thus a new $c_g$ dependence in the 
equations of motion.  Together, these results make the $c$- and 
$c_g$-dependence of the Shapiro time delay complicated and highly 
model-dependent.  The question of whether the time delay depends
on the speed of gravity is thus unanswerable: without a good deal 
of further physical input into the choice of a $c\ne c_g$ model, 
the dependence simply cannot be specified.

For the models I have studied here, the best existing observations 
restrict $\gamma$ to $|\gamma-1|<5\times10^{-5}$ \cite{Cassini}.
By (\ref{d14}), this requires $(1-2b)\epsilon<5\times10^{-5}$.
Note that this does not in itself restrict $c/c_g$; if $b$ is
close enough to $1/2$, $\epsilon$ can be large.  Within the
present framework, though---and given the observed restrictions
on $\gamma$, and thus the ``closeness'' to standard general 
relativity---it can be shown that the Fomalont-Kopeikin measurement
is mainly sensitive to the logarithmic term in (\ref{d13}).%
\footnote{This is unintuitive, since one expects a logarithm 
to be less sensitive than a power.  But VLBI measures differences
in arrival times, thus essentially differentiating a logarithm 
whose argument is small.  The relevant computation in the present
model is identical to that within general relativity, for which
details can be found in section 4 of \cite{Kopeikin3}; the 
velocity-dependent part of the argument of the logarithm gives 
the last term in equation (12) of that paper, which is large because 
of the smallness of the angle $\theta$ separating Jupiter and the 
quasar.  Note that because $c_g>c$ \cite{Moore}, $c_2$ in (\ref{d15}) 
must be either greater than $c$ (for $b\approx 1/2$) or approximately
equal to $c$ (for $\epsilon\approx 0$), so the prefactor cannot 
become much larger than in standard general relativity.}
As discussed above, to lowest order in 
$v$ this term depends only on the speed of light $c$.  One can 
thus argue that within the particular class of theories I have 
investigated, present measurements are not yet accurate enough to 
determine $c_g$.  

On the other hand, this analysis has also made it clear that even
this weaker claim depends on the class of ``comparison theories'' 
under consideration.  It is not easy to see how to change the
$c$-dependence of the argument of the logarithm in (\ref{d13});
the absence of $c_g$ to first order in $v$ follows from very
general properties of retarded potentials.  But it is not at all
clear that this term will always be the most important one.  For 
example, it is evident from (\ref{d12}) that one can choose 
$\epsilon$ and $\beta$ in such a way that $1+\zeta$ becomes very 
large, thus allowing a $c_g$-dependent prefactor in (\ref{d13}) to 
dominate over the logarithmic term.  For the simple models considered 
here, such a choice would make $\gamma$ unacceptably large.  But 
by adding additional terms to the Lagrangian (\ref{b4}) for $u$, 
one can easily introduce new couplings that can allow $\gamma$ and 
$1+\zeta$ to vary more independently. 

The particular models I have considered here may be testable by other 
observations.  In particular, they have preferred frame effects that 
might, for example, lead to a prediction of anomalous tides \cite{Will2}.  
This is a generic feature of models with $c\ne c_g$: there is a unique 
frame in which light and gravity both propagate isotropically, and 
motion with respect to this frame is, in principle, observable.  On 
the other hand, even in the general class of theories considered
here, I have examined only a narrow set.  The general kinetic term for 
$u$ has four coupling constants, three fundamental propagation speeds,
and dispersion relations that make these speeds energy-dependent
\cite{Jac2}, and it might be possible to tune the couplings to make 
preferred frame effects small while maintaining a nontrivial expression 
for Shapiro time delay.  This is an interesting question, but it lies 
outside the scope of this paper.

\vspace{1.5ex}
\begin{flushleft}
\large\bf Acknowledgments
\end{flushleft}

This work was supported in part by Department of Energy grant
DE-FG02-91ER40674.  I would like to thank Dave Mattingly for many
useful suggestions, and to acknowledge useful comments from Sergei
Kopeikin, Cliff Will, and Jean-Philippe Uzan.


\begin{thebibliography}{99}
\bibitem{Fomalont}  E.\ B.\ Fomalont and S.\ M.\ Kopeikin, Astrophys.\ J.\ 
  598 (2003) 704
\bibitem{Asada} H.\ Asada, Astrophys.\ J.\ 574 (2002) L69.
\bibitem{Will} C.\ M.\ Will, Astrophys.\ J.\ 590 (2003) 683.
\bibitem{Kopeikin0} S.\ M.\ Kopeikin, Phys.\ Lett.\ A312 (2003) 147.
\bibitem{Kopeikin} S.\ M.\ Kopeikin, Class.\ Quant.\ Grav.\ 21 (2004) 1.
\bibitem{Kopeikin2} E.\ B.\ Fomalont and S.\ M.\ Kopeikin, preprint 
  astro-ph/0311063.
\bibitem{Kopeikin3} S.\ M.\ Kopeikin, Astrophys.\ J.\ 556 (2001) L1.
\bibitem{Faber} J.\ A.\ Faber, preprint astro-ph/0303346.
\bibitem{Frittelli} S.\ Frittelli, Mon.\ Not.\ Roy.\ Astron.\ Soc.\
  344 (2003) L85.
\bibitem{Low} R.\ J.\ Low, Class.\ Quant.\ Grav.\ 16 (1999) 543.
\bibitem{Damour} T.\ Damour, in \emph{300 Years of Gravitation},
  edited by S.\ W.\ Hawking and W.\ Israel (Cambridge University Press, 1987),
  p.\ 128.
\bibitem{Jac} T.\ Jacobson and D.\ Mattingly, Phys.\ Rev.\ D64 (2001) 024028.
\bibitem{Kost} V.\ A.\ Kosteleck{\'y} and S.\ Samuel, Phys.\ Rev.\ D40
  (1989) 1886.
\bibitem{Uzan} G.\ F.\ R.\ Ellis and J.-P.\ Uzan, preprint gr-qc/0305099.
\bibitem{Lor} See, for example, \emph{CPT and Lorentz Symmetry},
  edited by V.\ A.\ Kostelecky (World Scientific, 1999).
\bibitem{Moore} G.\ D.\ Moore and A.\ E.\ Nelson, JHEP 0109 (2001) 023.
\bibitem{Jac2} T.\ Jacobson and D.\ Mattingly, preprint gr-qc/0402005.
\bibitem{MTW} C.\ W.\ Misner, K.\ S.\ Thorne, and J.\ A.\ Wheeler,
  {\em Gravitation} (W.\ H.\ Freeman, 1973), \S22.5. 
\bibitem{Carlip} S.\ Carlip, Phys.\ Lett.\ A267 (2000) 81.
\bibitem{Feynman} R.\ Feynman, R.\ B.\ Leighton, and M.\ L.\ Sands, \emph{The 
  Feynman Lectures on Physics}, Volume II, Chapter 21 (Addison-Wesley, 1989). 
\bibitem{Cassini} B.\ Bertotti, L.\ Iess, and P.\ Tortora, Nature 425 (2003) 
  374.
\bibitem{Will2} C.\ M.\ Will, \emph{Theory and experiment in gravitational
  physics} (Cambridge University Press, 1993), \S8.2.
\end{thebibliography}
\end{document}